\documentclass[useAMS,usenatbib,usegraphicx]{mn2e}

\title[OPserver]{OPserver: interactive online-computations of
opacities and radiative accelerations}
\author[C. Mendoza et al.]
{C. Mendoza,$^{1,2}$\thanks{E-mail: claudio@ivic.ve.}
M. J. Seaton,$^{3}$
P. Buerger,$^{4}$
A. Bellor\'{\i}n,$^{5}$
\newauthor
M. Mel\'endez,$^{6}$\thanks{Present address: Institute for
Astrophysics and Computational Sciences, Department of Physics,
The Catholic University of America, Washington, DC 20064, and
Exploration of the Universe Division, Code 667, NASA Goddard Space
Flight Center, Greenbelt, MD 20771, USA.}
J. Gonz\'alez,$^{1,7}$
L. S. Rodr\'{\i}guez,$^{8}$
F. Delahaye,$^{9}$
\newauthor
E. Palacios,$^{7}$
A. K. Pradhan$^{10}$
and C. J. Zeippen$^{9}$
\\
$^{1}$Centro de F\'{\i}sica, Instituto Venezolano de Investigaciones Cient\'{\i}ficas (IVIC),
      PO Box 21827, Caracas 1020A, Venezuela \\
$^{2}$Centro Nacional de C\'alculo Cient\'{\i}fico Universidad de Los Andes (CeCalCULA),
M\'erida 5101, Venezuela \\
$^{3}$Department of Physics and Astronomy, University College London, London WC1E 6BT, UK \\
$^{4}$Ohio Supercomputer Center, Columbus, Ohio 43212, USA \\
$^{5}$Escuela de F\'{\i}sica, Facultad de Ciencias, Universidad Central de Venezuela,
      PO Box 20513, Caracas 1020-A, Venezuela \\
$^{6}$Departamento de F\'{\i}sica, Universidad Sim\'on Bol\'{\i}var, PO Box 89000,
      Caracas 1080-A, Venezuela \\
$^{7}$Escuela de Computaci\'on, Facultad de Ciencia y Tecnolog\'{\i}a, Universidad de Carabobo,
       Valencia, Venezuela \\
$^{8}$Centro de Qu\'{\i}mica, Instituto Venezolano de Investigaciones Cient\'{\i}ficas (IVIC),
      P.O. Box 21827, Caracas 1020A, Venezuela \\
$^{9}$LUTH, Observatoire de Paris, F-92195 Meudon, France \\
$^{10}$Department of Astronomy, The Ohio State University, Columbus, Ohio 43210, USA
}
\begin{document}

\date{Accepted. Received; in original form }

\pagerange{\pageref{firstpage}--\pageref{lastpage}} \pubyear{2006}

\maketitle

\label{firstpage}

\begin{abstract}
Codes to compute mean opacities and radiative accelerations for arbitrary chemical
mixtures using the Opacity Project recently revised data have been restructured
in a client--server architecture and transcribed as a subroutine library. This
implementation increases efficiency in stellar modelling where element stratification
due to diffusion processes is depth dependent, and thus requires repeated fast opacity
reestimates. Three user modes are provided to fit different computing environments,
namely a web browser, a local workstation and a distributed grid.
\end{abstract}

\begin{keywords}
atomic processes -- radiative transfer -- stars: interior.
\end{keywords}


\section{Introduction}
Astrophysical opacities from the Opacity Project (OP) have been
updated by \citet{bad05} to include inner-shell contributions and
an improved frequency mesh. The complete data set of monochromatic
opacities and a suite of codes to compute mean opacities and
radiative accelerations (OPCD\_2.1\footnote{\tt
http://cdsweb.u-strasbg.fr/topbase/op.html}) have also been
publicly released by \citet{sea05} to make in-house calculations
for arbitrary mixtures more versatile and expedient. Regarding
data accuracy, there is excellent overall agreement between the
OPAL \citep{igl96} and OP results as discussed by \citet{sea04},
\citet{bad05} and \citet{del05}.

Rosseland mean opacities are sensitive to both the basic atomic
data used and the assumed abundances of the chemical elements. What had
been a good agreement between theory and the helioseismological
data was found to be less good using revised solar abundances from
\citet{asp05}. The revised OP opacities have been instrumental in
discussions of that problem \citep{ant05, bah05a, bah05b, bah05c,
bah05, del06}.

The modelling of stellar interiors, on the other hand, is being
renewed with the solar experience. Present ({\em WIRE}, {\em MOST},
{\em CoRoT}) and future ({\em Kepler}) space probes and the well
established solar methods are giving the field of asteroseismology
remarkable growth and the guarantee of invaluable data
\citep{met04, kur05, chr06}. In future work on stellar models it
may be desirable to take account of revisions in abundances
similar to those performed for the Sun.

For some types of stars, models must take into account
microscopic diffusion processes, e.g. radiative levitation,
gravitational settling and thermal diffusion, as they can affect
the internal and thermal structures, the depth of the convection
zone, pulsations and give rise to surface abundance anomalies
\citep{sea99, del05, bou06}. As reviewed by \citet{mic04}, such
processes are relevant in the description of chemically peculiar
stars, horizontal-branch stars, white dwarfs and neutron stars,
and in globular cluster age determinations from Population~II
turnoff stars. Furthermore, in order to solve the outstanding
discrepancy of the atmospheric Li abundance in old stars with that
predicted in big-bang nucleosynthesis, \citet{ric05} have proposed
Li sinking deep into the star due to diffusion. This hypothesis has been
recently confirmed in the observations by \citet{kor06}.

The OPCD\_2.1 release includes data and codes to compute the radiative
accelerations required for studies of diffusion processes. It
should be noted that the radiative accelerations are summed over
ionization stages and that data for the calculation of diffusion
coefficients are calculated assuming that the distribution over
ionization stages of the diffusing ions is the same as that in the
ambient plasma. The validity of this approximation is discussed by
\citet{GLAM}.

In some cases, particularly when element stratification depends on
stellar depth, calculations of mean opacities and radiative
accelerations must be repeated at each depth point of the model
and at each time step of the evolution, and thus the use of codes
more efficient than those in OPCD\_2.1 may be necessary. This
becomes critical in the new distributed computing grid
environments where the network transfer of large volumes of data
is a key issue. In the present work we have looked into these
problems, and, as a solution, report on the implementation of a
general purpose, interactive server for astrophysical opacities
referred to as {\tt OPserver}. It has been installed at the Ohio
Supercomputer Center, Columbus, Ohio, USA, from where it can be
accessed through a web page\footnote{\tt http://opacities.osc.edu}
or a linkable subroutine library. It can also be downloaded
locally to be run on a stand-alone basis but it will demand
greater computational facilities. In Section~2 we discuss the
computational strategy of the codes in OPCD\_2.1 followed by a
description of {\tt OPserver} in Section~3. In Section~4 we
include some tests as an indication of its performance with a
final summary in Section~5.


\section{OPCD codes}

We highlight here some of the key features of the codes in
OPCD\_2.1. For a chemical mixture specified by abundance fractions
$f_k$, they essentially compute two types of data: Rosseland mean
opacities (RMO) and radiative accelerations (RA).


\subsection{Rosseland mean opacities}

For the frequency variable
\begin{equation}
u=h\nu/k_{\rm B}T
\end{equation}
where $k_{\rm B}$ is the Boltzmann constant, RMO are given by the
harmonic mean of the opacity cross section $\sigma(u)$ of the
mixture
\begin{equation}
\frac{1}{\kappa_{\rm R}}=\mu\int_0^{v_\infty}\frac{1}{\sigma(u)}\ {\rm d}v
\end{equation}
where $\mu$ is the mean atomic weight. The $\sigma(u)$ is a
weighted sum of the monochromatic opacity cross sections for each
of the chemical constituents
\begin{equation}\label{sum}
\sigma(u)=\sum_k f_k \sigma_k(u) \ ,
\end{equation}
and is conveniently tabulated on the $v$-mesh
\begin{equation}
v(u) = \int_0^u \frac{F(u)}{1-\exp(-u)}\ {\rm d}u
\end{equation}
where
\begin{equation}
F(u)=\frac{15u^4\exp(-u)}{4\pi^4[1-\exp(-u)]^2}
\end{equation}
and $v_\infty=v(u \rightarrow \infty)$. The rationale behind the $v$-mesh
is that it enhances frequency resolution where $F(u)$ is large \citep{bad05}.


\subsection{Radiative accelerations}

Similarly, the RA for a selected $k$ element can be expressed as
\begin{equation}
g_{\rm rad} = \frac{\mu\kappa_{\rm R}\gamma_k}{c\mu_k}{\cal F}
\end{equation}
where $\mu_k$ is its atomic weight and $c$ the speed of light. The
function ${\cal F}$ is given in terms of the effective temperature
$T_{\rm eff}$ and fractional depth $r/R_\star$ of the star by
\begin{equation}
{\cal F} = \pi B(T_{\rm eff})(R_\star/r)^2
\end{equation}
with
\begin{equation}
B(T)=\frac{2(\pi k_{\rm B}T)^4}{15c^2h^3} \ .
\end{equation}
The dimensionless parameter
\begin{equation}
\gamma_k = \int \frac{\sigma_k^{\rm mta}}{\sigma}\ {\rm d}v
\end{equation}
depends on the cross section for momentum transfer to the $k$
element
\begin{equation}
\sigma_k^{\rm mta} = \sigma_k(u)[1-\exp(-u)] - a_k(u)
\end{equation}
where $a_k(u)$ is a correction to remove the contributions
of electron scattering and momentum transfer to the electrons.
Both $\sigma_k(u)$ and $a_k(u)$, which are hereafter referred to
as the {\tt mono} data set ($\sim$1~GB), are tabulated in
equally spaced $v$ intervals to facilitate accurate interpolation
schemes.


\begin{figure*}
http://vizier.u-strasbg.fr/topbase/opserver/fig1.eps
\caption{Flowcharts for the computations of Rosseland mean
opacities (RMO) and radiative accelerations (RA) with the codes in
the OPCD\_2.1 release. They are carried out in two stages: in a
time consuming Stage~1, data are computed for the whole $(T,N_e)$
plane followed by fast bicubic interpolations in Stage~2. The
intermediate files {\tt mixv.xx} and {\tt acc.xx} enable
communication between these two steps.}
\end{figure*}


\subsection{Computational strategy}

The computational strategy adopted in the OPCD\_2.1 release is depicted in the
flowcharts of Figure~1 where it may be seen that calculations of RMO
and RA are carried in two stages. In a time consuming Stage~1, RMO
and RA are computed with the {\tt mixv} and {\tt accv} codes,
respectively, on a representative tabulation of the complete
temperature--electron-density $(T,N_e)$ plane.
In {\tt mixv} the chemical mixture is specified in the
input file {\tt mixv.in} as
\begin{equation}\label{mix1}
\{X,Z,N,Z_k,f_k\}
\end{equation}
where $X$ and $Z$ are the hydrogen and metal mass-fractions, $N$
the number of elements, and $Z_k$ and $f_k$ are the metal nuclear charges
and fractional abundances.
In {\tt accv}, the input data ({\tt accv.in}) are
\begin{equation}\label{mix2}
\{N,Z_k,f_k,Z_i,N_\chi,\chi_j\}
\end{equation}
where now $k$ runs over the $N$ elements of the mixture, and
$Z_i$ and $\chi_j$ are respectively the nuclear charge and $N_\chi$
abundance multipliers of the test $i$ element. Input data formats in either
{\tt mixv.in} or {\tt accv.in} give the user flexible control
over chemical mixture specifications.

As shown in Figure~1, the intermediate output files {\tt mixv.xx}
($\sim$85 KB) containing
\begin{equation}
\{\rho,\kappa_{\rm R}\}(T,N_e)\ ,
\end{equation}
where $\rho$ is the mass-density, and {\tt acc.xx} ($\sim$470 KB)
with
\begin{equation}
\left\{\kappa_{\rm R}, \frac{\partial\kappa_{\rm R}}{\partial\chi},
\gamma, \frac{\partial\gamma}{\partial\chi}\right\}(T, N_e, \chi_j)
\end{equation}
are written to disk. They are then respectively read by the codes {\tt
opfit} and {\tt accfit} in Stage~2 for fast bicubic
interpolations of RMO and RA on stellar depth profiles
$\{T,\rho,r/R_\star\}(i)$ specified by the user in the
{\tt opfit.in} and {\tt accfit.in} input files. The final
output files are {\tt opfit.xx} containing
\begin{equation}
\left\{\log \kappa_{\rm R}, \frac{\partial \log\kappa_{\rm
R}}{\partial\log T}, \frac{\partial\log \kappa_{\rm R}}{\partial\log \rho}\right\}(i)
\end{equation}
and {\tt accfit.xx} with
\begin{equation}
\{\log \kappa_{\rm R}, \log \gamma, \log g_{\rm rad}\}(i,\chi_j)\ .
\end{equation}

In this computational approach, performance is mainly limited by
the summation in equation~(\ref{sum}) which implies disk reading
the {\tt mono} data set; for instance, in {\tt mixv} it takes up
to $\sim$90\% of the total elapsed time. OPCD\_2.1 also includes other
codes such as {\tt mx} and {\tt ax} which respectively compute RMO
and RA for a star depth profile. The chemical mixture can be fully varied
at each depth point using the specifications in equations (\ref{mix1}--\ref{mix2}),
the RMO and RA being obtained in a one-step process using bicubic
interpolations without splines. These methods are thus suitable for cases
with multi-mixture depth profiles \citep{sea05}. Further details of all the OPCD
codes are contained in the reference
manual\footnote{\tt http://opacities.osc.edu/publi/OPCD.pdf}.


\begin{figure*}

http://vizier.u-strasbg.fr/topbase/opserver/fig2.eps
\caption{OPserver enterprise showing the web-server--supercomputer
tandem at the Ohio Supercomputer Center (OSC) and the three
available user modes. (A) The {\tt OPlibrary} and monochromatic
opacities ({\tt mono}) are downloaded locally and linked to the
user modelling code such that RMO/RA are computed locally. (B) The
{\tt OPlibrary} is downloaded locally and linked to the modelling
code but RMO/RA are computed remotely at the OSC. (C) RMO/RA
computations at the OSC through a web client.}
\end{figure*}


\section{OPserver}

In {\tt OPserver} the computational capabilities of the codes in
OPCD\_2.1 are greatly enhanced by the following innovative
adaptations.

\begin{enumerate}

\item The codes are restructured within a client--server network
architecture whereby the time consuming Stage~1 is performed on a
powerful processor while the fast Stage~2 is moved to the client,
e.g. a web server or a user workstation. In this arrangement
performance could be affected by the client--server transfer of
the {\tt mixv.xx} and {\tt acc.xx} intermediate files, but since
they are never larger than 0.5 MB, it is not expected to be a
deterrent with present-day bandwidths. In a local installation
where both the client and server reside on the same machine,
communication is managed through shared buffers in main memory; i.e.
the corresponding data in {\tt mixv.xx} and {\tt acc.xx} are not
written to disk.

\item The codes are transcribed as a subroutine library---to be
referred to hereafter as the {\tt OPlibrary}---which can be linked
by the user stellar modelling code for recurrent subroutine calls
that avoid data writing on disk. That is, the input data
in the {\tt mixv.in}, {\tt accv.in}, {\tt opfit.in} and {\tt accfit.in}
files and the output tables in the {\tt opfit.xx} and {\tt accfit.xx}
files (see Figure~1) are now handled as subroutine parameters while
the intermediate {\tt mixv.xx} and {\tt acc.xx} files are passed via
shared main-memory buffers. Chemical mixtures are again specified with
the formats of equations (\ref{mix1}--\ref{mix2}) which allow
full variation at each depth point in a single subroutine call.

\item RMO/RA are computed with the complete {\tt mono} data set always
loaded in main memory thus avoiding lengthy and repeated disk readings.
This is achieved by implementing {\tt OPserver} on a dedicated server
where {\tt mono} is permanently resident in RAM, or in the case of
a local installation, by disk-reading it once at the outset of a
modelling calculation.

\item When accessing the remote server, client data requests are
addressed through the HTTP protocol, i.e. in terms of a Uniform
Resource Locator (URL). This allows data fetching from the central
facility through an interactive web page or a network access
subroutine, the latter being particularly suitable for a stellar
model code that is to be run in a distributed grid environment.

\item The {\tt do-loop} that computes the summation of
equation~(\ref{sum}) has been parallelized in OpenMP which
provides a simple, scalable and portable scheme for shared-memory
platforms.

\end{enumerate}

As shown in Figure~2, the current OPserver enterprise is implemented as a
client--server model at the Ohio Supercomputer Center (OSC).  The web server
communicates with the supercomputer via a socket interface.  Earlier versions
were developed on an SGI Origin2000 server with the PowerFortran parallelizing
compiler. The current version runs on a Linux system with Fortran OpenMP directives.
{\tt OPserver} offers three user modes with full functionality except when otherwise
indicated in the following description.

\begin{description}

\item[{\bf Mode~A}] In this mode {\tt OPserver} is set up locally
on a stand-alone basis (see Figure~2).  The facilities of the OSC are not used. A
new OPCD release (OPCD\_3.3\footnote{\tt
http://cdsweb.u-strasbg.fr/topbase/op.html}) is downloaded, followed
by (i) installation of both the {\tt OPlibrary} and the {\tt mono}
data set and (ii) linking of the {\tt OPlibrary} to the user
modelling code. Computations of RMO/RA are preceded by the
reading of the complete {\tt mono} data set from disk and
therefore requires at least 1~GB of RAM.

\item[{\bf Mode~B}] In this mode, the {\tt OPlibrary} is downloaded, installed
and linked to the user code, but Stage~1 is performed remotely at
the OSC (see Figure~2). This option has been customized for stellar modelling in
a distributed grid environment that would otherwise imply (i.e.
Mode~A) the network transfer, installation and disk-reading of the
{\tt mono} data set at runtime. It is also practical when local
computer capabilities (RAM and/or disk space) are limited. The functions provided by
the {\tt mx} and {\tt ax} codes have not been implemented.

\item[{\bf Mode~C}] In this mode RMO/RA computations at the OSC are requested
through an interactive web page\footnote{\tt http://opacities.osc.edu} which
allows both Stage~1 and Stage~2 to be carried out remotely or,
alternatively, Stage~2 locally by downloading the {\tt
mixv.xx/acc.xx} intermediate files (see Figure.~1) with the browser
for further processing with local {\tt opfit}/{\tt accfit} executables.

\end{description}

\section{Tests}

{\tt OPserver} benchmarks were initially carried out on an SGI Origin2000
multiprocessor at the OSC with an earlier release of OPCD. For the standard
S92 mixture \citep{sea94}, the {\tt mixv} code took up to 140~s to compute
the {\tt mixv.xx} file, of which 126~s were dedicated to disk-reading and
14~s to the actual computing of the mean opacities. {\tt OPserver} took on
average $12.0\pm 0.5$~s to compute {\tt mixv.xx} which was not written to disk
unless requested. In Figure~3 we show the acceleration obtained on the Origin2000
through parallelization where the calculation of mean opacities is reduced to
2~s with 8 processors. Further significant acceleration is prevented by data
transfer overheads.

On more recent workstations, the local performances of the codes in OPCD\_2.1 and
{\tt OPserver} depend on processor speed and RAM and cache sizes. For instance,
on a PowerMac G5 (PowerPC 970fx processor at 2.0 GHz, 1GB of RAM and L2 cache of 512 KB)
the first time {\tt mixv} is run it takes for a single S92 mixture
103.8~s to compute the RMO, but on subsequent runs the elapsed time is reduced to
an average of $28.2\pm 0.2$~s. Similarly, {\tt OPserver} takes 103.3~s which is then
reduced to $31.4\pm 0.4$~s on subsequent runs. Once the {\tt mono} data set is loaded
in RAM by {\tt OPserver} (Mode A), calculations of RMO for a single S92 mixture only take
$5.29\pm 0.02$~s and $6.13\pm 0.01$~s for RA for the test element Ar. In Mode~B, where
Stage 1 is carried out remotely at the OSC and the {\tt mixv.xx} and {\tt accv.xx} files
are transferred at the relatively low rate of 1.88 KB/s, computations of RMO and RA
take $5.9\pm 0.1$~s and $9.3\pm 0.3$~s, respectively. The noticeable longer time taken
for the latter is due to the transfer time taken for the larger {\tt accv.xx} file.

\section{Summary}
Rosseland mean opacities and radiative accelerations can be computed from OP data
in any one of the following ways.
(i) Download the original OPCD\_2.1 package as described by \citet{sea05} and
perform all calculations locally.

\noindent
(ii) Mode A, download the upgraded OPCD\_3.3 package, install {\tt OPserver}
and perform all calculations locally by linking the subroutines in the {\tt OPlibrary}.
Calculations with {\tt OPserver} are more efficient but require large local computer
memory.

\noindent
(iii) Mode B, as Mode A but with Stage 1 performed remotely at the OSC. Mode B is
convenient if fast calculations are required but local computer
memory is limited or when stellar modelling is to be carried out in a grid environment.

\noindent
(iv) Mode C, perform all calculations remotely at the OSC through an interactive web page
 whereby files are downloaded locally with the browser.


\section*{Acknowledgments}
We acknowledge the invaluable assistance of Juan Luis Chaves and Gilberto D\'{\i}az
of CeCalCULA during the initial stages of {\tt OPserver}. We are also much indebted
to the Ohio Supercomputer Center, Columbus, Ohio, USA, for hosting {\tt OPserver}
and for technical assistance; to the Centre de Donn\'ees  astronomiques
de Strasbourg, France, for hosting the OPCD releases; and to Drs Josslen Aray, Manuel
Bautista, Juan Murgich and Fernando Ruette of IVIC for allowing us to test the OPserver
installation on different platforms. FD would like to thank S.~Rouchy for technical support.
AKP and FD have been partly supported by a grant from the US National Science Foundation.


\begin{figure}
\includegraphics[width=84mm]{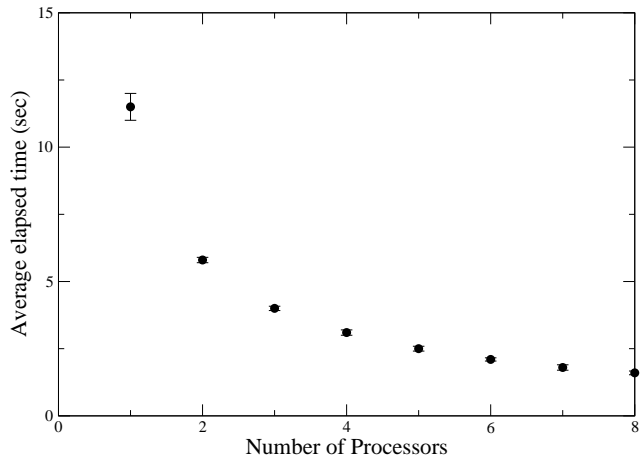}
\caption{Average elapsed time for the computation of RMO (Stage~1)
in {\tt OPserver} as a function of the number of processors on a
SGI Origin2000 showing the acceleration obtained through
parallelism. The corresponding time taken by the {\tt mixv} code for this
calculation is 140~s where 126~s are taken by data reading from disk.}
\end{figure}

\bsp

\label{lastpage}

\end{document}